# Risk of Harm in VR Dating from the Perspective of Women and LGBTQIA+ Stakeholders

DEVIN TEBBE*, MERYEM BARKALLAH*, and BRAEDEN BURGER*, University of Michigan-Flint, USA
DOUGLAS ZYTKO, University of Michigan-Flint, USA

Virtual reality (VR) dating introduces novel opportunities for romantic interactions, but it also raises concerns about new harms that typically occur separately in traditional dating apps and general-purpose social VR environments. Given the subjectivity in which VR dating experiences can be considered harmful it is imperative to involve user stakeholders in anticipating harms and formulating preventative designs. Towards this goal with conducted participatory design workshops with 17 stakeholders identified as women and/or LGBTQIA+; demographics that are at elevated risk of harm in online dating and social VR. Findings reveal that participants are concerned with two categories of harm in VR dating: those that occur through the transition of interaction across virtual and physical modalities, and harms stemming from expectations of sexual interaction in VR.

## 1 INTRODUCTION

Virtual reality (VR) dating has emerged as a novel avenue for individuals seeking romantic and sexual connections [3]. Unlike conventional dating apps, VR dating offers immersive experiences where users interact with avatars and simulate physical proximity. However, this unique landscape also presents new risks through the convergence of harms that commonly occur in general social VR such as harassment [1] and mobile dating apps such as physical sexual violence [4]. Towards designing safe VR dating experiences we are involving demographics at disproportionate risk of harm in social VR and online dating - women and LGBTQIA+ individuals - in anticipating harms and crafting mitigative solutions. In this workshop paper we outline new potential harms unique to VR dating from their perspective.

## 2 METHOD

We held a series of participatory design sessions with 17 women and LGBTQIA+ stakeholders, organized into 5 groups that each engaged in 6+ hours of discussion and design activities across three sessions. The first two sessions used consent in VR [2] as a lens to differentiate acceptable and unwanted behavior in VR dating and how VR daters should interact to avoid committing nonconsensual behavior. Participants articulated their viewpoints through scenarios which were used as the basis for producing and reflecting on technologies for safety in sessions 2-3.

---

*All three authors contributed equally to this research.

## 3 FINDINGS

Participants anticipated several risks of harm, or nonconsensual behavior, in VR dating that can serve as focal points for future design of safety mechanics. We focus on two below.

**Transitional harms:** Participants focused on harms that might occur when a user takes a relationship from the virtual world to the physical world. VR serves as a space where users can mold their visual and social self-presentation, which increases the potential for users to establish a false impression of a partner's physical-world self. Participants expressed concern that some users might establish a false understanding of a partner's boundaries through interactions in the virtual world, and then inadvertently violate their boundaries when in the physical world. P7 demonstrates this with hand-holding: *"...if you touch in VR, or if you like, held hands in VR, and you just assume now that we're meeting in public [in the physical world], we can also hold hands."* Retaliatory harm resulting from this impression violation is described by Alexandra: *"Say you go out on a date in real life, and you're not interested anymore. And then what if the other person's response was really like violent, very, like, physically threatening in response to that."*

**Harm stemming from expectations of romantic/sexual interaction:** Many participants were worried that the denial of sexual interactions in VR dating may result in retaliatory harm. Furthermore, participants anticipated that some users might unintentionally cause harm by assuming consent to an interaction when consent was not explicitly given, additionally noting that assumed consent can be especially prevalent in VR dating where body language cannot be easily expressed. Interestingly, some participants discussed that consent within VR dating may be considered trivial or even unnecessary to some users due to its detachment from the physical world which is unique even from other forms of online dating, as shown by Alexandra in a discussion about consenting to a sexual interaction: *"Because they're like, well, it's not it's not real, right? Like, why not [engage in the behavior without consent]?"*

## 4 AUTHOR BIOS

**Devin Adam Tebbe** and **Braeden Burger** are undergraduate students at the University of Michigan-Flint majoring in Computer Science. Their research interests center on VR consent and the unique harms that occur within XR environments. **Meryem Barkallah** is a research scholar at University of Michigan-Flint in the USA and a Computer Science Engineering student at Ecole Polytechnique de Sousse in Tunisia. Her current research involves developing VR/AR mechanics to mitigate nonconsensual behavior. **Douglas Zytko** is an Associate Professor at the University of Michigan-Flint. His NSF-funded research uses computer-mediated consent as a lens for understanding and designing preventative technologies for sexual violence.

## ACKNOWLEDGMENTS

This work is partially supported by the U.S. National Science Foundation under Grant No. IIS-2211896.